# Surface optical Raman modes in InN nanostructures

Satyaprakash Sahoo,[1] M. S. Hu,[2] C. W. Hsu,[3] C. T. Wu,[3] K. H. Chen,[2,3] L. C. Chen,[3] A. K. Arora[1] and S. Dhara[1,4,a)]

[1] Materials Science Division, Indira Gandhi Centre for Atomic Research, Kalpakkam-603102, India

[2] Institute of Atomic and Molecular Sciences, Academia Sinica, Taipei-106, Taiwan

[3] Center for Condensed Matter Sciences, National Taiwan University, Taipei-106, Taiwan

[4] Department of Electrical Engineering, Institute for Innovations and Advanced Studies, National Cheng Kung University, Tainan-701, Taiwan.

## Abstract

Raman spectroscopic investigations are carried out on one-dimensional nanostructures of InN, such as nanowires and nanobelts synthesized by chemical vapor deposition. In addition to the optical phonons allowed by symmetry; $A_1$, $E_1$ and $E_2$(high) modes, two additional Raman peaks are observed around 528 cm$^{-1}$ and 560 cm$^{-1}$ for these nanostructures. Calculations for the frequencies of surface optical (SO) phonon modes in InN nanostructures yield values close to those of the new Raman modes. A possible reason for large intensities for SO modes in these nanostructures is also discussed.

[a)] Author to whom correspondence should be addressed; electronic mail: dhara@igcar.gov.in

Among III-nitride semiconducting compounds, InN has attracted a lot of attention to many researchers in the recent years for its promising application in optical and high performance electronic devices.[1-5] Due to high drift velocity at room temperature; InN is a suitable candidate for field effect transistor.[6] InN has been explored as a solar cell material with high efficiency.[7] It has the possibility of band gap tuning in InAlGaN covering the range of 0.7–6 eV for various optoelectronic applications including radiation damage resistant solar cell applications.[3] Vibrational properties of III-nitride nanocrystals are also interesting, as the study of phonon leads to the understanding of the basic electron-phonon coupling and electron intraband transitions in these materials. The wurtzite crystal structure of InN belongs to $C_{6v}$ point group symmetry. A group symmetrical analysis shows the following six allowed optical modes; $A_1$, $2B_1$, $E_1$, $2E_2$. Out of which $A_1$ and $E_1$ are both Raman and IR active, $E_2$ mode is only Raman active and $B_1$ mode is inactive both in Raman and IR. In wurtzite InN, the polar $A_1$ and $E_1$ modes split into longitudinal optic (LO) and transverse optic (TO) components with different frequencies due to macroscopic electric fields associated with the LO phonons. As a result a total of six Raman active modes are reported,[8] which are $A_1$ (LO), $A_1$ (TO), $E_1$ (LO), $E_1$ (TO), $E_2$(high) and $E_2$(low) at 586, 447, 593, 476, 488 and 87 cm$^{-1}$ respectively (Table 1).

Several Raman studies have been carried out by various researchers both on bulk and in nanostructured InN.[2,8,9] In polar crystals, surface optic modes have been observed on the low frequency side of LO mode. When the size of the crystal reduces to few nanometers, changes are observed in the phonon spectrum. In particular, when the phonon is confined to the surface, its frequency lies between TO and LO frequencies.[10] The dielectric constant $\varepsilon(\omega)$ is negative between the LO and TO mode. The SO phonon frequency and intensity mainly depend on the size and the shape of the nanostructured material. Recently, large intensities in surface optical phonons have been reported in polar semiconductor nanowires [11,12] and single GaN nanowire.[13] High intensities of SO modes for the large surface area available to the nanostructure may be quite useful in detecting small diameter oscillations in the reduced dimension.[12] There are hardly any report of surface optical (SO) phonon in the nanostructured InN system. In a recent study, Chang e*t al.* reported SO phonon at 543 cm$^{-1}$ in 2μm thick epitaxial InN film by coherent phonon spectroscopy.[14] Here we report results of Raman spectroscopic investigation of 1-D nanostructures (nanowires and nanobelts) of InN. The calculated frequencies for the wurtzite structure are compared with those of the observed additional Raman peaks.

InN nanostructres were grown on Si (100) substrates using chemical vapour deposition. Metallic In was used for nanowires (NW) and metal-organic trimethylindium was used for the nanobelt (NB) as precursors. Au was used as catalyst in the vapour-liquid-solid (VLS) process and ammonia was used the reactant gas. Details of the synthesis of NW, and NB samples are given else where.[15,16] The morphology and structural details of InN nanostructures were investigated by field emission scanning electron microscopy (FESEM; JEOL 6700) and high resolution transmission electron microscopy (HRTEM, JEM-4000EX). Raman scattering measurements were performed using 532 nm line of a diode-pumped solid state laser as the excitation and analyzed using a monochromator (Jobin-Yvon U1000), equipped with liquid nitrogen cooled CCD detector for recording the spectra.

Fig. 1 (a) shows the FESEM image of InN NW with a length in the range of 10-20 μm and diameter ranging from 75 to 150 nm [high magnification image, inset in Fig. 1(a)]. Low magnification image of the belts like morphology with a length ranging from 20 to 60 µm are observed [Fig. 1 (b)] for NB sample. High magnification image [inset in Fig. 1(b)] shows width of these NB are in the range of 40-200 nm and thickness from 50 to 100 nm. HRTEM image [Fig. 2(a)] shows *d*-spacing (~ 0.308 nm) corresponding to (100) planes of wurtzite InN NW. Selected area electron diffraction [SAED, inset in Fig. 2 (a)] confirms the wurtzite InN phase recorded with a [001] zone axis. The TEM study reveals that growth of the NW samples is in the [110] direction. Typical bright field TEM image of the NB sample is shown in Fig. 2(b). The dark stripes in low magnification image of the NB result from bending contours, which are normally found in bent thin crystals. The observed interplanar spacing of the lattice planes is 0.308 nm, which corresponds to the (100) lattice planes of wurtzite InN. Growth of NB samples are along the [110] direction. The inset of Fig. 2b shows a typical SAED pattern with zone axes lying along [001] of the wurtzite phase.

A typical Raman spectrum of NW and NB samples are shown in Fig. 3(a) and (b), respectively. Distinct peaks at about ~ 447 and 490 cm$^{-1}$ along with broad spectra ranging from 500 to 600 cm$^{-1}$ are observed for both the samples. However, distinct shoulders at different frequencies with different intensities are seen in the broad spectra. In order to resolve these peaks a peak fit program was used and the resolved peaks are also shown in Fig. 3. The positions of various peaks identified in these all different nanostructures and their assignments are mentioned in the table I. These assignments are done by comparing the observed mode frequencies with those of reported Raman data for InN.[8] It may be pointed out that the fitted profile matches well with the measured spectra. A comparison of the Raman peak positions reported for InN film with those found in the present study (Table I) shows that $A_1$(LO) and $E_1$(LO) modes appear at slightly lower frequencies, than those of bulk InN. This red shift of phonon frequencies is generally attributed to phonon confinement effect.[17] When the size of the crystal reduces to very small length scale, of the order of few nanometer, the lattice periodicity is interrupted at the particle surface which leads to relaxation of $q = 0$ selection rule, where $q$ is the scattering vector. Thus the phonon wave function has to decay to a small value close to the boundary. This allows all the $q$'s in the Brillouin zone to contribute to Raman intensity and results in shift as well as broadening of Raman peak in nanocrystals.

Apart from the five Raman active modes expected between 400 to 650 cm$^{-1}$, two additional modes appear around 528 cm$^{-1}$ and 560 cm$^{-1}$. The mode around 528 cm$^{-1}$ can not be attributed to Si substrate as the bulk Si peak occurs at 520 cm$^{-1}$ and is much narrower than the 528 cm$^{-1}$ peak. This was confirmed by recording the Raman spectrum of the substrate. Davydov *et al.* observe a peak at 561 cm$^{-1}$ in several wurtzite InN samples.[8] They suggest that this peak may be due to the participation of phonon other than zone centre phonon in scattering process. In another report, Qian *et al.* argue that the peak at 561 cm$^{-1}$ is associated with disordered in InN.[9] In view of these ambiguous assignments of the mode around 560 cm$^{-1}$, we examine the possibility of surface phonon being responsible for the new modes found for InN NW and NBs.

The SO phonon frequency and intensity mainly depend on the size and the shape of the nanostructured material. Now we calculate the SO phonon modes that associate with all the samples of InN. The expression for SO phonon for a 1-D nanostructure is given by,[10]

$$\omega^2{}_{so} = \omega^2{}_{TO} \frac{\varepsilon_0 - \rho_{nx}\varepsilon_m}{\varepsilon_\infty - \rho_{nx}\varepsilon_m} \qquad (1)$$

Where $\omega_{TO}$ is the frequency of TO phonon, $\varepsilon_0$ and $\varepsilon_\infty$ and are the static and high frequency dielectric constant of the material and $\varepsilon_m$ is the dielectric constant of the medium. $\rho_{nx}$ is given by

$$\rho_{nx} = \frac{K_1(x)I_0(x)}{I_1(x)K_0(x)} \qquad (2)$$

where $I_n$, and $K_n$ are the modified Bessel functions and $x=qr$ ($r$ being radius of the nanostructure). Here, we use 10.3 and 6.7 for the values of $\varepsilon_0$ and $\varepsilon_\infty$, respectively for InN.[18] The dielectric constant of the (air) medium is taken as 1. As there are two polar phonons $A_1$ and $E_1$ in the wurtzite structure, one could in principle expect SO phonons associated with each of the polar phonons. Here we have obtained both the surface optical phonons. Fig. 4 shows the $\omega_{SO}$ as a function of $qr$. If one considers $q$ to be the same as the scattering vector corresponding the back scattering geometry ($q=2k=4\pi/\lambda_{Ex}$) for $\lambda_{Ex}$ = 532 nm wavelength, the value of $qr$ lies in a range that is determine by the spread in diameters of the nanowires. Typically for 100 nm diameter, $qr$ can be calculated as 1.18. Fig. 4 also shows the frequencies of the new modes and the range of $qr$ applicable to the present experiments. Note that the observed mode frequencies lie very close to the expected SO a phonon frequency curves. Based on the good agreement one may assign the modes around 528 cm$^{-1}$ and 560 cm$^{-1}$ to SO modes with $A_1$ and $E_1$ characters, respectively. The broad nature of the SO phonon essentially arises due to the wide range of the NW and NB diameter/dimension. One can see that the intensities of these SO phonons are comparatively high. One of the reasons is that a considerable fraction of atoms in nanowire reside on the surface than the interior of the wire. Furthermore, rough surfaces/interfaces are known to yield stronger interface/surface phonons.[12,19] From the present micrographs one can clearly see the diameter fluctuation and consequent rough surfaces of the nanostructures. This makes the intensity of the surface mode comparable to that of the other phonons.

Raman spectroscopic investigation of InN nanostructures such as nanowire and nanobelts, shows the appearance of two distinct new peaks around 528 cm$^{-1}$ and 560 cm$^{-1}$. Calculation pertaining to surface optic phonon modes in InN reveals two SO phonons whose frequencies are close to the observed new modes. High intensities of SO modes arise due to large surface area and the surface roughness associated with the nanostructure.

**Figure Captions:**

Fig.1. Low magnification FESEM images of randomly oriented InN nanostructures of a) nanowires and b) nanobelts. Insets show the high magnification images of the corresponding nanostructures.

Fig.2. Bright field TEM images for a) nanowires and b) nanobelts. Structural studies of lattice imaging shows wurtzite InN with growth direction along [110]. Insets show the SAED pattern corresponding to the nanostructures with zone axes recorded along [001].

Fig.3. Raman spectra for InN nanostructures of a) nanowires and b) nanobelts. Peaks arising due to different modes are resolved by fitting the spectra to a set of seven peaks. Full curve: total fitted spectrum, dashed curves: individual fitted peaks. Peaks corresponding to surface optic mode are indicated by arrows.

Fig.4. Calculated SO phonon frequencies as a function of $qr$, full curve: SO($E_1$), dashed curve: SO($A_1$), horizontal full and dashed lines are the TO and LO frequencies of $E_1$ and $A_1$ modes respectively. Symbols are measured SO frequencies, open symbol: NW, filled symbol: NB.

Table 1: Raman modes in nanowires and nanobelt samples of wurtzite InN

| Raman modes | Calculated (cm$^{-1}$) | Nanowire (cm$^{-1}$) | Nanobelt (cm$^{-1}$) |
|---|---|---|---|
| $A_1$(TO) | 447[a] | 449 | 447 |
| $E_1$(TO) | 476[a] | 478 | 476 |
| $E_2$(high) | 488[a] | 490 | 492 |
| SO ($A_1$) | 527[b] | 528 | 527 |
| SO ($E_1$) | 560[b] | 563 | 556 |
| $A_1$(LO) | 586[a] | 584 | 575 |
| $E_1$(LO) | 593[a] | 590 | 588 |

[a] Ref. 8 for Bulk, [b] Present work for 1-D nanostructure.

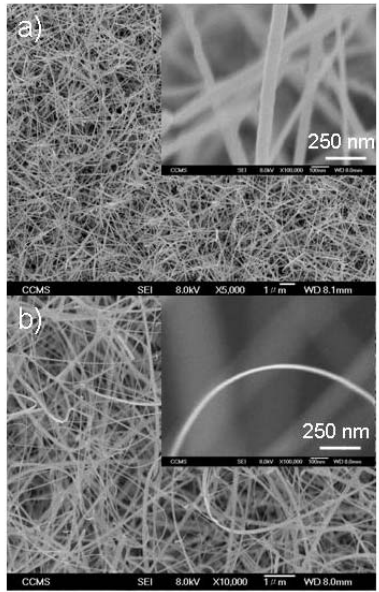

Fig. 1

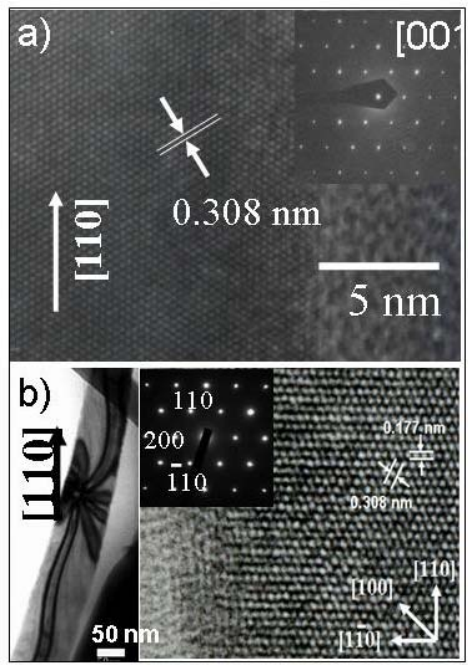

Fig. 2

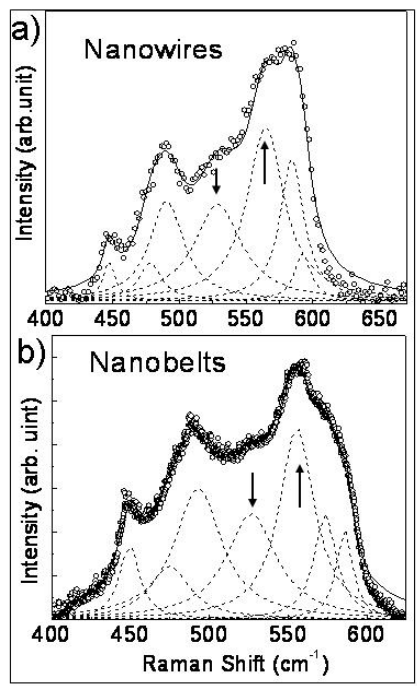

Fig. 3

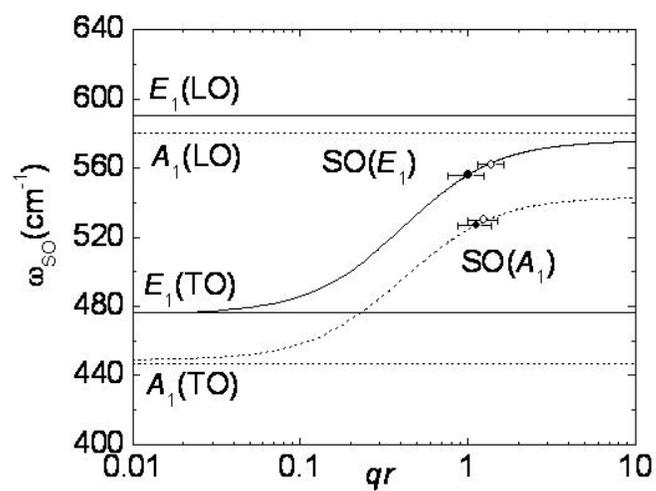

Fig. 4